\documentclass[10pt]{IEEEtran}

\usepackage{amsmath,amssymb}
\usepackage{color}
\usepackage[pdftex]{graphicx}
\usepackage{epstopdf}
\usepackage{epsfig}
\usepackage{bm}
\usepackage{cancel}
\usepackage{algorithm}
\usepackage{algorithmic}
\usepackage{hyperref}

\DeclareSymbolFont{AMSb}{U}{msb}{m}{n}
\DeclareMathSymbol{\R}{\mathalpha}{AMSb}{"52}

\begin{document}

\title{An optimal transport model for imaging in atmospheric turbulence}

\author{J. M. Nichols$^a$, A. Watnik$^a$, T. Doster$^a$, S. Park $^b$, A. Kanaev$^a$, L. Cattell$^b$, and G. K. Rohde$^b$\\
$^a$ Naval Research Laboratory, Optical Sciences\\
$^b$ University of Virginia, Dept. of Electrical and Computer Engineering, Dept. of Biomedical Engineering
}

\IEEEtitleabstractindextext{
\begin{abstract}

We describe a new model for image propagation through open air in the presence of changes in the index of refraction (e.g. due to turbulence) using the theory of optimal transport. We describe the relationship between photon density, or image intensity, and the phase of the traveling wave and, together with a least action principle, suggest a method for approximately recovering the solution of the photon flow. By linking atmospheric propagation solutions to optimal transport, we provide a physics-based (as opposed to phenomenological) model for predicting turbulence-induced changes to sequences of images. Simulated and real data are utilized to validate and compare the model to other existing methods typically used to model this type of data. Given its superior performance in describing experimental data, the new model suggests new algorithms for a variety of atmospheric imaging applications. 
\end{abstract}

}

\maketitle 

\IEEEdisplaynontitleabstractindextext

\section{Introduction \label{sec:intro}}

Atmospheric turbulence has long been a source of distortion in open air imaging applications.  Spatial and temporal fluctuations in the physical properties of the atmosphere (e.g., temperature, humidity) give rise to variability in the index of refraction, thereby altering the optical signal.  In imaging applications, the end result is degraded image or video data while for free space optical communications, the turbulence corrupts the link causing a higher bit error rate.  Efforts to mitigate these errors have been hindered to a large extent by the lack of practical, accurate models for the solution of the wave equation in the presence of atmospheric turbulence. 

Here we demonstrate a new solution based on minimization of kinetic energy using optimal transport.  The resulting transport model is efficient to compute, invertible, and can be estimated from easily obtained intensity measurements (i.e. images).  Moreover, the model is not phenomenological (e.g., convolution \cite{Bertozzi:13}, optical flow \cite{Mitzel:09}) but is shown to be consistent with  the physics associated with the image formation.  For this reason, we hypothesize the transport-based approach to image modeling might offer improved predictions of imagery collected in a turbulent medium.  Indeed, the model is demonstrated here to provide a more accurate, parsimonious model of sequences of turbulence-corrupted imagery than does optical flow \cite{Mitzel:09}.  The solution has potentially important implications for any application involving propagation of an electro-magnetic field through a medium with varying refractive index.

\section{Atmospheric propagation as a Transport Problem \label{sec:Maxwell}}

The goal of this section is to describe the propagation of an electromagnetic (EM) field through the atmosphere as a transport problem.  As will be shown, transport models are consistent with the problem physics and admit practical, computational solutions.  

The starting point for the study of propagating EM radiation is Maxwell's equations for isotropic materials \cite{Ryan:91}
\begin{subequations}
\begin{align}
\nabla\times{\bf E}({\bf x})&=i\omega \mu_0 {\bf H}({\bf x}) 
\label{eqn:max1}\\
\nabla\times {\bf H}({\bf x})&=-i\omega \epsilon_0{\bm\epsilon}({\bf x}){\bf E}({\bf x})
\label{eqn:max2}\\
\mu_0\nabla\cdot {\bf H}({\bf x})&=0
\label{eqn:max3}\\
\epsilon_0\nabla\cdot \left({\bm\epsilon}({\bf x}) {\bf E}({\bf x})\right)&=0
\label{eqn:max4}
\end{align}
\end{subequations}
where ${\bf E}({\bf x})$ is the electric field intensity vector in (V/m), ${\bf H}({\bf x})$ is the magnetic field intensity vector in (A/m), ${\bf B}({\bf x}) = \mu_0 {\bf H}({\bf x})$ is the magnetic field induction vector in (Wb/m) and ${\bf D}({\bf x}) = {\bm\epsilon({\bf x})}{\bf E}({\bf x})$ is the electric field displacement vector in (C/m) and $\omega$.  The radiation is assumed to be mono-chromatic, with time dependence governed by the angular frequency $\omega$ \cite{Ryan:91}.  The vector ${\bf x}$ specifies the full 3-dimensional space ${\bf x}\equiv (x_1,x_2,z)$, where $z$ is the direction of propagation.

The quantity ${\bm\epsilon}({\bf x})$ is the relative complex permittivity of the atmosphere while the constants $\epsilon_0,~\mu_0$ are the vacuum dielectric constant and free space (vacuum) permeability respectively.  Note also that in forming Eqn. (\ref{eqn:max3}), it is assumed that the relative permeability of the atmosphere is unity which allows us to further relate the relative complex permittivity to the complex index of refraction via \cite{Ryan:91}
\begin{align}
{\bm\epsilon}({\bf x})&\equiv \left[n({\bf x})+i\kappa({\bf x})\right]^2
\label{eqn:index}
\end{align}
where $n({\bf x})$ is the usual refractive index and $\kappa({\bf x})$ is referred to as the extinction coefficient.  In what follows is assumed that the latter is negligible so that we may write ${\bm\epsilon}({\bf x})=n({\bf x})^2$

Taking the curl of Eqn. (\ref{eqn:max1}) and then substituting in Eqn. (\ref{eqn:max2}-\ref{eqn:max4}) yields the vector wave equation
\begin{align}
\nabla^2{\bf E}({\bf x})+\nabla\left({\bf E}({\bf x})\cdot\frac{\nabla {\bm\epsilon}({\bf x})}{\bm\epsilon({\bf x})}\right)+k_0^2{\bm\epsilon}({\bf x}){\bf E}({\bf x})&=0
\label{eqn:Helmholtz}
\end{align}
where $k_0=\sqrt{\epsilon_0\mu_0}\omega=1/\lambda_0$ is the wavenumber and $\lambda_0$ the associated wavelength.  The second term in this expression is a direct result of applying the constitutive relationship, Eqn. (\ref{eqn:max4}), giving
\begin{align}
\nabla\cdot {\bf E}({\bf x})&=-{\bf E}({\bf x})\cdot \frac{\nabla {\bm\epsilon}({\bf x})}{{\bm\epsilon}({\bf x})}.
\label{eqn:constraint}
\end{align}
However, this term is typically neglected as it is assumed that either the atmosphere is homogeneous, or that the relative permittivity is nearly unity ($\frac{\nabla {\bm\epsilon}({\bf x})}{{\bm\epsilon}({\bf x})}=\nabla\log({\bm\epsilon}({\bf x}))\approx 0$ if ${\bm\epsilon}({\bf x})\approx 1$).  Indeed, we will show later that if one considers the constitutive equation (\ref{eqn:constraint}), the resulting contribution to the transport-based framework is higher-order in terms of the turbulence-induced perturbations to the refractive index. 


\subsection{Transforming the Parabolic Wave Equation}

Leaving out the second term in (\ref{eqn:Helmholtz}), the parabolic wave equation can be derived by replacing the (vector) electric field with the scalar field
\begin{align}
{\bf E}({\bf x})=\Psi(\vec{x},z)e^{ik_0 z}
\label{eqn:wave}
\end{align}
where $\vec{x}=(x_1,x_2)$ defines the plane in the direction transverse to propagation.  This representation assumes a wave propagating horizontally (in the $\hat{z}$ direction) in air  with wavenumber $k_0$.  Note that in making this substitution we are replacing a vector with a complex scalar.  This substitution (scalar for a vector) is mathematically justified, since the Laplacian operator in (\ref{eqn:Helmholtz}) is separable in terms of the field components. More importantly, Eqn. (\ref{eqn:wave}) is justified on physical grounds by noting that for a propagating EM plane wave, the electric field vector is confined to the transverse plane (negligable polarization in the ``$z$'' direction).  The complex scalar amplitude $\Psi(\vec{x},z)$ is therefore sufficient to capture both the magnitude and polarization direction (i.e., phase angle in the transverse plane associated with real and imaginary parts) of the electric field (see \cite{Saleh:91}, section 5.4).  Note also that had we not assumed a negligible extinction coefficient there would be a real portion of the exponent in (\ref{eqn:wave}) governing the decay of the solution.  In short, for the application of interest, the vector-to-scalar wavefield transformation is both mathematically convenient and physically meaningful (see e.g., \cite{Born:99}, section 8.4).

Substituting (\ref{eqn:wave}) into (\ref{eqn:Helmholtz}) gives 
\begin{align}
i2k_0\frac{\partial{\Psi}(\vec{x},z)}{\partial z}+\nabla_{X}^2\Psi(\vec{x},z)+k_0^2\eta(\vec{x},z)\Psi(\vec{x},z)=0.
\label{eqn:parabolic}
\end{align}
where the operator $\nabla_X^2$ denotes the Laplacian operating in the two transverse coordinates and $\eta(\vec{x},z)\equiv n^2(\vec{x},z)-1$ is the deviation in refractive index from unity.  Additionally, we have neglected dispersion as is commonly done, i.e. $|\partial_{zz}\Psi(\vec{x},z)|<<2k_0|\partial_z{\Psi}(\vec{x},z)|$. 

It is important to note that this expression possesses a strong similarity to the Schr\"{o}dinger equation where the last term in (\ref{eqn:parabolic}) plays the role of a potential function \cite{Flatte:86}.
Based on this similarity, one can pursue similar analysis techniques.  Here, we use the so-called Madelung transformation \cite{Madelung:27,Tsang:06,Vadasz:16} (also known as the Luneberg-Kline transformation \cite{Ott:97}) and represent the field as $\Psi(\vec{x},z)=\sqrt{\rho(\vec{x},z)}\exp(i\phi(\vec{x},z)/2)$ where it is assumed $\rho(\vec{x},z)\ge 0$.  Combined with appropriate re-scaling of the spatial coordinates (see Appendix \ref{sec:appendixA}), Eqn. (\ref{eqn:parabolic}) becomes
%
\begin{subequations}
\begin{align}
&\frac{\partial  \rho(\vec{x},z)}{\partial z}+\nabla_X\cdot \left(\rho(\vec{x},z) v(\vec{x},z)\vphantom{A_1^2}\right)=0
\label{eqn:continuity2}\\
&\frac{\partial v(\vec{x},z)}{\partial z}+(v(\vec{x},z)\cdot\nabla_X)v(\vec{x},z)=2\nabla_X \gamma(\eta(\vec{x},z)).
\label{eqn:momentum2}
\end{align}
\end{subequations}
where $v(\vec{x},z)\equiv \nabla_X \phi(\vec{x},z)$ and the function
\begin{align}
\gamma(\eta(\vec{x},z))&\equiv -\nabla^2_X\log(n^2(\vec{x},z))+(\nabla_X\log(n^2(\vec{x},z)))^2\nonumber \\
&~~~~~~~+\eta(\vec{x},z)
\label{eqn:eta}
\end{align}
is solely a function of the refractive index.  The first two terms in (\ref{eqn:eta}) arise due to the ``diffraction term'' \cite{Gureyev:95} (alternatively the ``quantum potential'' \cite{Bohm:84}), which naturally appears as $\frac{\nabla_X^2(\rho(\vec{x},z)^{1/2})}{\rho(\vec{x},z)^{1/2}}$ in (\ref{eqn:momentum2}), but can be re-cast in terms of the refractive index using the constitutive relationship (\ref{eqn:constraint}) (see Appendix \ref{sec:appendixB}).

 {\it Thus the parabolic wave equation can be readily interpreted as the familiar continuity and momentum equations from fluid mechanics where the phase gradient $v(\vec{x},z)=\nabla_X \phi(\vec{x},z)$ plays the role of the velocity, the ``density'' $\rho(\vec{x},z)=\Psi(\vec{x},z)\Psi(\vec{x},z)^*$ is the image intensity, and the refractive index creates the potential function $2\gamma(\eta(\vec{x},z))$.}
 
Now note that Eqn. (\ref{eqn:momentum2}) could also be written solely in terms of the phase variable (see Appendix \ref{sec:appendixA}) as the familiar Hamilton-Jacobi equation, or in fluid mechanics terminology, the unsteady Bernoulli equation
\begin{align}
\frac{\partial\phi(\vec{x},z)}{\partial z}+\frac{1}{2}(\nabla_X\phi(\vec{x},z))^2=2\gamma(\eta(\vec{x},z)).
\label{eqn:momentum3}
\end{align}
Moreover, for small perturbations to the index $\eta\ll 1$ the approximation $\log(1+\delta)\approx \delta$ for $\delta\ll 1$ means we could alternatively have written $\gamma(\eta(\vec{x},z))\approx -\nabla^2_X\eta(\vec{x},z)+(\nabla_X\eta(\vec{x},z))^2+\eta(\vec{x},z)$.  We can therefore neglect the first two terms so that $\gamma(\eta(\vec{x},z))\approx \eta(\vec{x},z)$.  Based on the genesis of these terms (discussion surrounding Eqn. \ref{eqn:eta}), this approximation is tantamount to the assumption that $\nabla^2_X\rho(\vec{x},z)^{1/2}/\rho(\vec{x},z)^{1/2}\ll 1$, one which is often made in optics \cite{Saleh:91},~\cite{Gureyev:95}. 

We will therefore seek an approach to modeling images that is consistent with the physics described by Eqns (\ref{eqn:continuity2},~\ref{eqn:momentum2} \&~\ref{eqn:momentum3}).  First, however, we briefly discuss some existing solutions.

\subsection{Prior art}

Some researchers have attempted to solve Eqn. (\ref{eqn:parabolic}) directly via numerical methods (see e.g., \cite{Kuttler:91}).  Such methods are known to be computationally intensive \cite{Wheeler:85}, thereby leading to approximate methods (see e.g., \cite{Leland:94}), or by instead focusing only on the statistical properties of the solution (see e.g., Fannjiang and Solna \cite{Fannjiang:05}).  None of these approaches are suitable for modeling sequences of images.  

The ``transport'' form of  Eqn. (\ref{eqn:parabolic}) has been leveraged by other research in optics, perhaps most notably as a means of phase retrieval under the heading of ``Transport Intensity Equation'' (TIE) approaches \cite{Gureyev:95},~\cite{Petruccelli:13}.  The focus in the TIE method is on (\ref{eqn:continuity2}) as it is assumed that intensity measurements are made over short propagation distances such that (\ref{eqn:momentum2}) can be ignored \cite{Petruccelli:13}, an assumption we cannot make in imaging.

Related applications have used the same basic Madelung transformation followed by the ``Wentzel-Kramers-Brillouin'' (WKB) approximation (high frequency approximation whereby one equates terms of common wavenumber) to analyze equations of the form (\ref{eqn:parabolic}) \cite{Benamou:99}.  In the context of the Schr\"{o}dinger equation, WKB analysis also yields the system of equations (\ref{eqn:continuity2}) and (\ref{eqn:momentum3}) (see e.g. \cite{Liu:06} \cite{Jin:11}).   A common solution is the method of characteristics, a Lagrangian approach that numerically integrates the spatial coordinates of the phase front (e.g., rays) forward in time (see e.g., \cite{Nazarathy:82},~\cite{Moussa:03}).  The main challenges are the problem size (each ray is integrated separately), and the associated numerical errors \cite{Jin:11}.   Methods that rely on a fixed grid (so-called Eulerian methods), can overcome the problem size and resolution issues, but tend to suffer from multi-valued solutions arising due to the nonlinearity in (\ref{eqn:momentum3}) \cite{Liu:06} which require other approximations and numerical procedures to alleviate \cite{Benamou:04}, (see also \cite{Blanc:00} and the references therein).  Inversion of these numerical methods is similarly challenging. 

Moreover, in the context of image propagation the WKB analysis is equivalent to geometric optics, where the first term in (\ref{eqn:momentum3}) is neglected \cite{Saleh:91}\cite{Benamou:99}.  
Thus, the WKB approximation does not actually solve the paraxial wave equation, a point that was recently highlighted by Potvin \cite{Potvin:15}.  In this work it is important to retain (and solve) the full expression, Eqn. (\ref{eqn:momentum3}), as this allows us to formally connect solutions of the parabolic wave equation to optimal transport theory in Section (\ref{sec:transport}).  

Due to the deficiencies of these physics-based models, the typical approach in image processing is to pursue phenomenological models that are practical, yet preserve certain features of the physical process.  To this end, by far the most popular approaches to modeling turbulence-corrupted images are convolution and optical flow; both have seen use in turbulence-mitigation.  A recent discussion of deconvolution methods applied to this problem can be found in \cite{Zhu:13} while an optical flow implementation of turbulence mitigation was explored in \cite{Mao:12}.  In section (\ref{sec:test}) we will, in fact, compare our physics-based model to an optical flow model in terms of their respective abilities to predict turbulence-corrupted images.  

In section (\ref{sec:transport}) we will derive a solution that is both practical {\it and} consistent with the problem physics by making the connection to optimal transport theory.    
By doing so, we can leverage the tremendous progress in optimal transport \cite{Gustavoreview:16} and develop a fast, accurate solution that works for very large problem sizes (e.g., Mega-pixel images), does not require time-marching, and is easily invertible (a pre-requisite for several applications).  

\subsection{Model Interpretation}

\begin{figure}[tb]
  \centerline{
   \begin{tabular}{c}
    \includegraphics[scale=0.375]{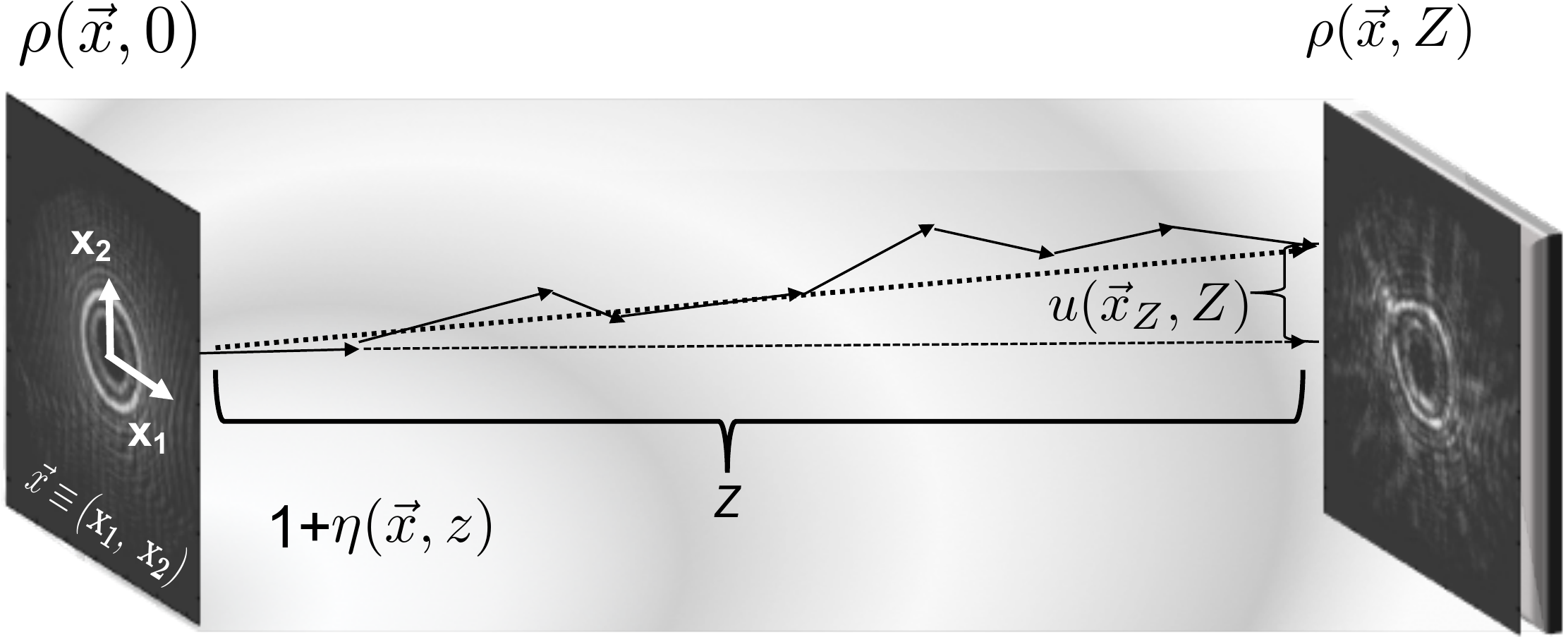}
  \end{tabular}  
  }
  \caption{Illustration of the transport problem.  Intensity is transported in the transverse plane as the associated EM field moves through space from $z=0$ to $z=Z$.  The transport model described here assumes the intensity is being transported along constant velocity paths, i.e., straight lines.  Each point on the source image is therefore mapped to a point on the corrupted image by a linear path.  The transverse displacement is denoted $u(\vec{x}_Z,Z)$; an expression for this displacement and its relation to the model (\ref{eqn:continuity2},~\ref{eqn:momentum2}) are given below.}
  \label{fig:scheme}
\end{figure}

Before proceeding to the solution, it is helpful to first consider the interpretation of the model (\ref{eqn:continuity2},~\ref{eqn:momentum2}).  Figure (\ref{fig:scheme}) depicts an example EM field propagating through an atmosphere governed by a varying index of refraction, quantified by the index perturbations $\eta(\vec{x},z)$. Note that the geometry of the wave propagation here allows us to view the $z$ dimension as time and thus we are able to exchange $z$ for $t$.

The structure of Eqns. (\ref{eqn:continuity2}-\ref{eqn:momentum3}) allows us to interpret the movement of an image through space as a transport problem that can be solved using recently developed tools (as will be shown in the next section).  The original image intensity $\rho(\vec{x},0)$ located at $\vec{x}$ is moved in directions defined by the phase gradient in the transverse plane.  The directions can be different at each transverse location and will change as $z$ (alternatively time) progresses.  The changes in direction are due to variations in the refractive index.
     
For example, in the absence of turbulence or other index fluctuations, the right hand side of Eqn. (\ref{eqn:momentum2}) disappears and the momentum equation becomes simply $Dv(\vec{x},z)/Dz=0$  where $D(\cdot)/Dz$ denotes the ``total derivative''.  Thus, in a homogeneous medium, and recalling the equivalence between $z$ and $t$, Eqn. (\ref{eqn:momentum2}) suggests there will be no transport in the transverse direction.  This makes sense as our (initially) paraxial rays are not experiencing refraction in this case, hence no intensity is being moved in the transverse plane.   Moreover, because the right-hand side is a function of the transverse index {\it gradient}, this statement also holds in the case that the refractive index is varying in $z$ only.  The phase will change with $z$ in this case (by Eqn. \ref{eqn:momentum3}), but the intensity will still move from source to destination in horizontal, straight lines (i.e., $Dv(\vec{x},z)/Dz$ is still 0).  

\begin{figure}[tb]
  \centerline{
   \begin{tabular}{c}
    \includegraphics[scale=0.375]{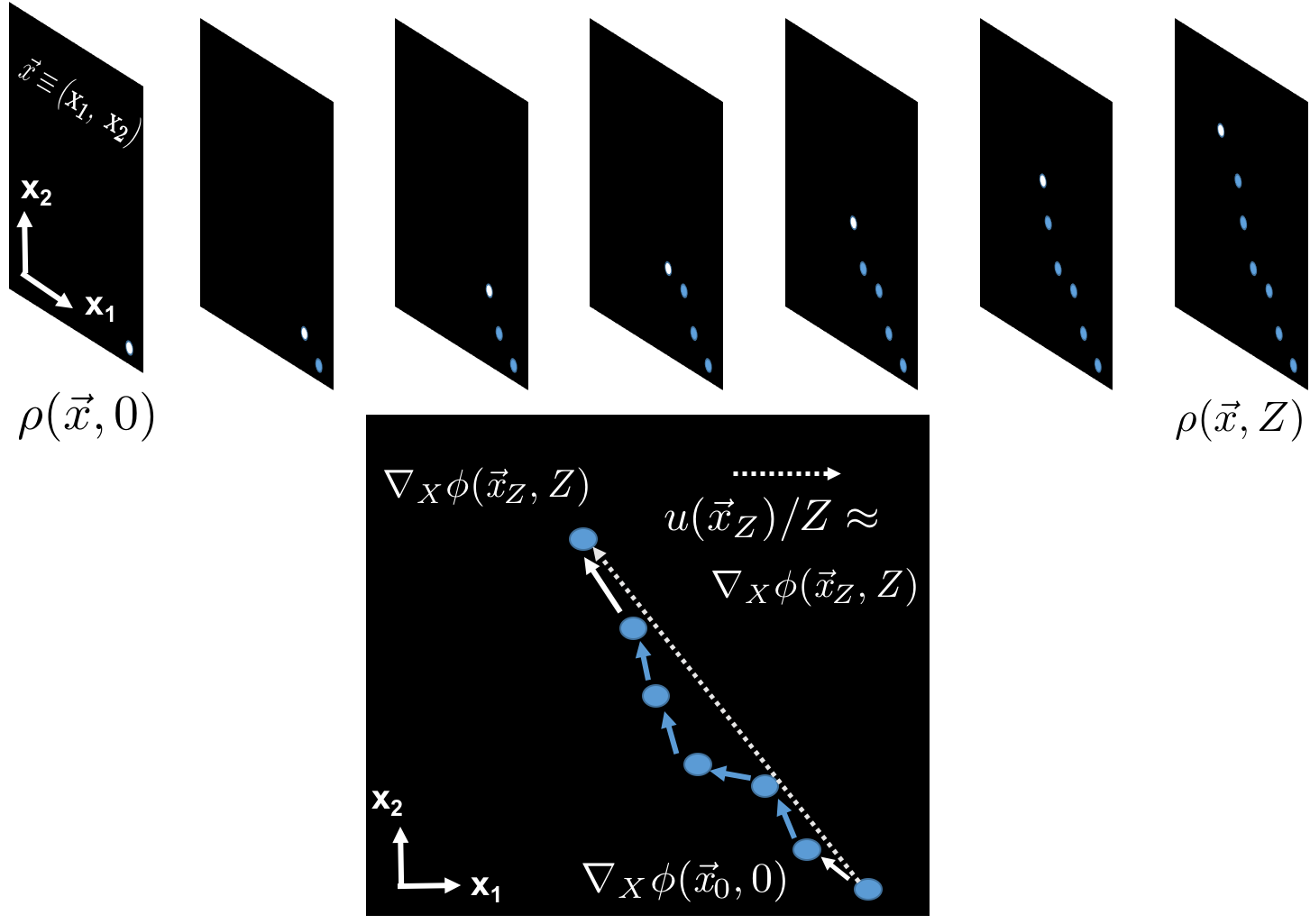}
  \end{tabular}  
  }
  \caption{In the transport modeling approach, one can think of the motion as occurring only in the transverse plane (lower plot) with the direction of propagation implicitly included as a time coordinate.  In this example, a single point is being perturbed by a series of step changes in refractive index.  Assuming only the first and last images are available, this approach is modeling the transport as constant velocity, linear motion between those two images.  The quality of this approximation will clearly depend on the strength of the index fluctuations and the distance $Z$ between the images used in creating the model.  As $Z\rightarrow 0$ or $\nabla_X \eta(\vec{x},z)=0$ the model is exact.  }
  \label{fig:path}
\end{figure}

Transport therefore occurs when a transverse index gradient causes refraction, at which point the intensity moves in the transverse plane along directions dictated by $\nabla_X\phi(\vec{x},z)$.  To illustrate, Figure (\ref{fig:path}) shows an image of a single point being transported in the transverse plane as time progresses.  The direction of propagation does not appear explicitly in the lower figure but rather is implicit in defining the transport path.  In this example, the index of refraction clearly possesses a series of steps in its transverse gradient, thereby causing the point to move in the transverse plane (absent such a gradient no apparent transverse motion would occur).  Assuming we can only observe the first and last images, we are using the constant velocity model $u(\vec{x}_Z,Z)/Z\approx v(\vec{x}_Z,Z)$ where $u(\vec{x}_Z,Z)$ denotes the displacement experienced by the point as it moves from location $\vec{x}_0$ to $\vec{x}_Z$.
As implied by the figure, this model will approach the true velocity as $Z\rightarrow 0$.  We now address the question of how to obtain the model from observed data.  

\section{Solutions via Optimal Transport \label{sec:transport}}

In this section we will demonstrate how to solve for both $\rho(\vec{x},z)$ and $v(\vec{x},z)$ for $z=0\cdots Z$ given a single pair of images $\rho(\vec{x},0),~\rho(\vec{x},Z)$ and absent information about the refractive index profile.  The solution is unique under the stated assumptions, computationally efficient and invertible, and can be estimated from intensity measurements (i.e., images) rendering it practically useful.    The resulting model can 1) be used to understand and predict the effects of turbulence on the imagery and 2) be inverted so that given an image, $\rho(\vec{x},Z)$, we can solve for $\rho(\vec{x},0)$.  

To see how, we first define the kinetic energy associated with moving image intensity over a distance $z=[0,Z]$ and corresponding time interval $t=[0,T]$
\begin{align}
\mathcal{A}\equiv Z\int_{\R^2}\int_0^Z \rho(\vec{x},z)|v(\vec{x},z)|^2dzd\vec{x}.
\label{eqn:action}
\end{align}
In continuum mechanics this quantity is often referred to as the {\it action} associated with a non-dissipative dynamical system without external forces or potentials \cite{Meirovitch:97}.  Now, of course, there is a potential function associated with this problem corresponding to the last term in (\ref{eqn:parabolic}) and given by $V(\vec{x},z)=2\gamma(\eta(\vec{x},z))$.  However, given the modest influence of the potential on the transport, recall $\eta(\vec{x},z)\ll 1$, we neglect this term in forming the action.  The consequences of this decision are discussed in what follows, along with results that justify this assumption (see Section \ref{sec:test}).

The principle of action minimization is a familiar one and has been used to derive the equations of motion for many dynamical systems, including Eqns. (\ref{eqn:continuity2},~\ref{eqn:momentum2}).  In fact, it has recently been shown that minimization of the specific action (\ref{eqn:action}) given the constraint (\ref{eqn:continuity2}) (intensity is conserved), yields precisely (\ref{eqn:momentum2}) along with the requirement that $v(\vec{x},z)=\nabla_X\phi(\vec{x},z)$ \cite{Dejan:16}, a relationship that came about naturally in our derivation of Eqn. (\ref{eqn:momentum2}).  It is therefore appropriate to study (\ref{eqn:action}) in formulating solutions to the parabolic wave equation (equivalently, Eqns \ref{eqn:continuity2} and \ref{eqn:momentum2}) for the case where index fluctuations are small.   
   
Making explicit the analogy between the system (\ref{eqn:continuity2},~\ref{eqn:momentum2}) and the associated action (\ref{eqn:action}) allows us to leverage ``optimal transport'' theory and the associated computational tools to solve for $\rho(\vec{x},z),~v(\vec{x},z)$.  The theory of optimal transport has in fact shown that there is only one solution to equation (\ref{eqn:continuity2}) that minimizes (\ref{eqn:action}) and possesses endpoints $\rho(\vec{x},0)$ and $\rho(\vec{x},Z)$ \cite{Villani:08},~\cite{Gustavoreview:16}.  

To develop this connection more fully, we take the Lagrangian perspective of the fluid system (\ref{eqn:continuity2},~\ref{eqn:momentum2}).  In this view the coordinates defining the transverse plane, $\vec{x}$, are no longer fixed, but change according to the system dynamics.  With this in mind, we label the coordinates over which the image is defined according to their location along the direction of propagation, e.g. $\vec{x}_z$ is the support of the image at $z$.  The dynamic coordinates are defined by the {\it Lagrangian flow map}, $\vec{x}_z\equiv f(\vec{x}_0,z)$ which evolves the starting coordinates $\vec{x}_0$ forward in space to location $z$.  This also means that $\dot{f}(\vec{x}_0,z)=v(f(\vec{x}_0,z),z)$ is the velocity \cite{Brenier:89}.

Returning to the continuity equation (\ref{eqn:continuity2}), we can see this is nothing more than a statement of total intensity conservation.  That is to say $\int \rho(\vec{x}_z,z)=\int \rho(\vec{x}_0,0)$.  This relationship can be re-written in terms of our previously defined mapping as
\begin{align}
\det(J_f(\vec{x}_0,z))\rho(\vec{x}_z,z)=\rho(\vec{x}_0,0)
\label{eqn:transform}
\end{align}
where $J_f(\vec{x}_0,z)$ denotes the Jacobian of $f(\vec{x}_0,z)$ (see \cite{Brenier:89}, \cite{Gustavoreview:16} or \cite{Dejan:16}) (note that in writing Eqn. \ref{eqn:transform} there is an implicit assumption that the coordinate transformation is smooth). Thus, knowledge of the Lagrangian flow map and its time rate of change are sufficient to define our solution.

Indeed, recent works have demonstrated that one can obtain the unique flow map so that the resulting intensity and velocity fields are consistent with minimization of (\ref{eqn:action}).  Specifically, it has been shown that the minimization 
\begin{align}
d_p(0,Z)^2&=\inf_{f}\int_{\R^2}\|f(\vec{x}_0,Z)-\vec{x}_0\|^2\rho(\vec{x}_0,0)d\vec{x}\nonumber \\
&=\min_{v}  \mathcal{A},
 \label{eqn:Kanto}
\end{align}
subject to the constraints imposed by the continuity equation (\ref{eqn:continuity2}), produces a coordinate transformation $f(\vec{x}_0,Z)$ that can be used to solve (\ref{eqn:momentum2}) \cite{Brenier:00},~\cite{Dejan:16}.    Note that the displacements being minimized, $u(\vec{x}_Z)\equiv f(\vec{x}_0,Z)-\vec{x}_0$, are in the transverse direction only.  

In deriving the relationship (\ref{eqn:Kanto}) it can also be shown that the minimizing solutions possess constant velocity which, in Lagrangian coordinates, is simply $u(\vec{x}_Z)/Z$.  Put another way, the turbulence-induced perturbations captured in the image pair $\rho(\vec{x},0),~\rho(\vec{x},Z)$ are modeled as growing linearly as the image moves from $z=0$ and $z=Z$.

This also means we can linearly interpolate the displacement coordinates  $f(\vec{x}_0,z)=(1-z/Z)\vec{x}_0+\frac{z}{Z} f(\vec{x}_0,Z)$ to obtain the image at {\it any} point in time via Eqn. (\ref{eqn:transform}).  This is consistent with our earlier assertion that, in the absence of index fluctuations, light moves in straight lines.  Finally, because the velocity (which is constant in $z$) must be expressed as a phase gradient \cite{Dejan:16}, we have
\begin{align}
v(\vec{x}_z,z)&=(f(\vec{x}_0,Z)-\vec{x}_0)/Z=\nabla_X\phi(\vec{x}_z,z)
\label{eqn:interp2}
\end{align}
thereby completing the solution to (\ref{eqn:continuity2},~\ref{eqn:momentum2}).  Note, the phase function in (\ref{eqn:interp2}) is the same as that used in defining the complex field amplitude in (\ref{eqn:parabolic}).   
{\it Provided that we accept the physical principle of action minimization we can indeed solve (\ref{eqn:continuity2},~\ref{eqn:momentum2}) and, by extension (\ref{eqn:parabolic}), given a single pair of clean/corrupted images and a means of solving (\ref{eqn:Kanto}).  The solution is the coordinate transformation $f(\vec{x}_0,z)$ from which we can obtain the image intensity via (\ref{eqn:transform}) and the velocity via (\ref{eqn:interp2})}.  This solution is exact if the index perturbations are zero; in the event that the index is fluctuating, the constant velocity solutions are approximating a wandering path with a straight line (see again Fig. \ref{fig:scheme} and Fig. \ref{fig:path}).  

What's more, as reviewed in \cite{Gustavoreview:16}, numerous numerical methods for solving (\ref{eqn:Kanto}) have emerged in recent years and are readily available. The model is simple to invert, handles very large problem sizes, does not require time-marching, and most importantly, is true to the physics of the problem.
 In the following section we will demonstrate the efficacy of this modeling approach and draw comparisons to traditional ``optical flow'' methods.

\section{Testing the Model \label{sec:test}}

In this section we test the applicability of the optimal transport-based model for imaging under turbulence developed above using both simulated and real imagery.  Because the model is consistent with the problem physics, we hypothesize it will perform well relative to phenomenological models. 

\subsection{Simulation}
To demonstrate the validity of our model, we verify whether Eqn. \eqref{eqn:interp2} holds in a simulated experiment. By verifying that under turbulence intensity travels in a straight path (constant velocity), we can indirectly verify whether optical flow solutions (all of which occur in straight paths) are compatible with the turbulence phenomenon.

 We consider an experiment whereby an image is passed through several ``phase screens'' in order to mimic the effects of the spatially varying refractive index \cite{Lane:92}.  
\begin{figure}[tb]
  \centerline{
   \begin{tabular}{ccc}
    \includegraphics[scale=0.25]{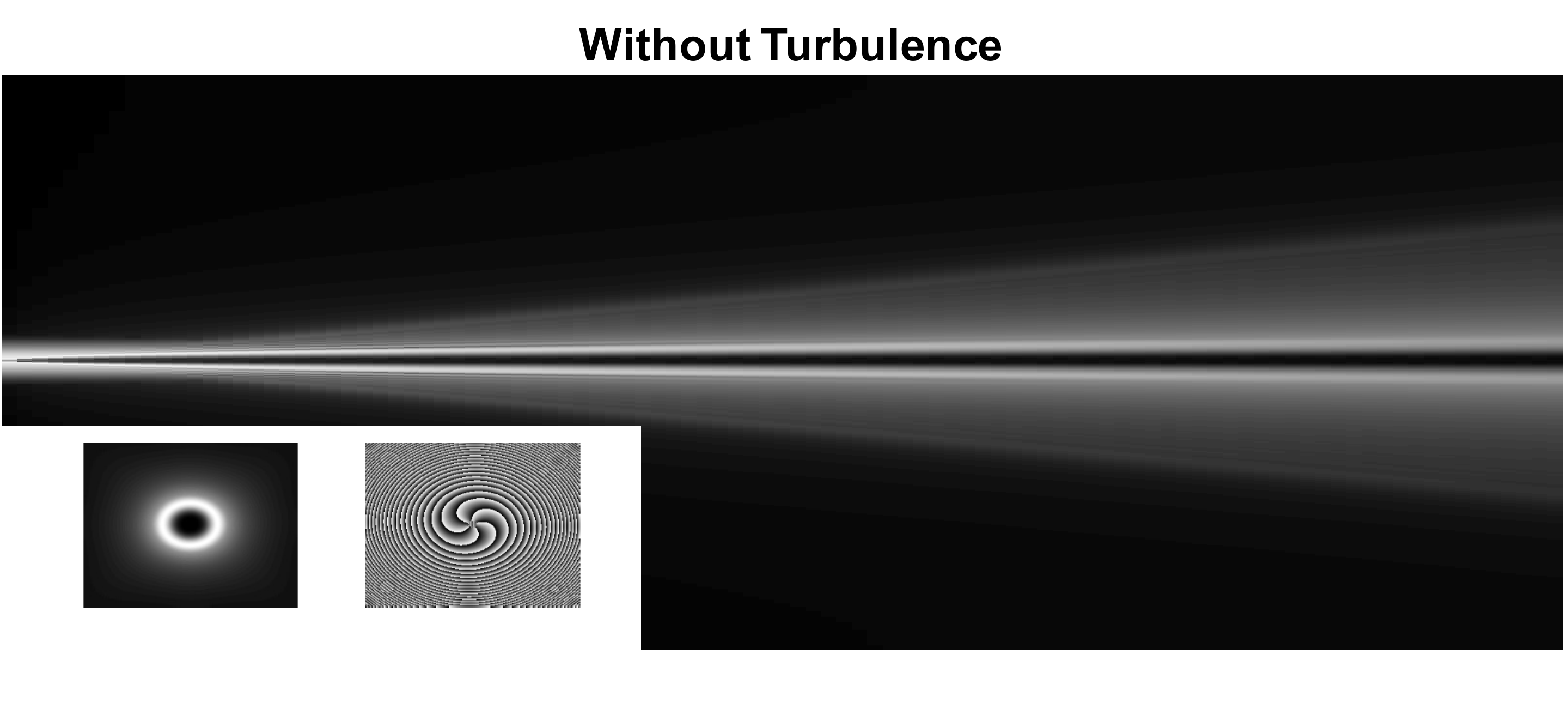} \\
     \includegraphics[scale=0.25]{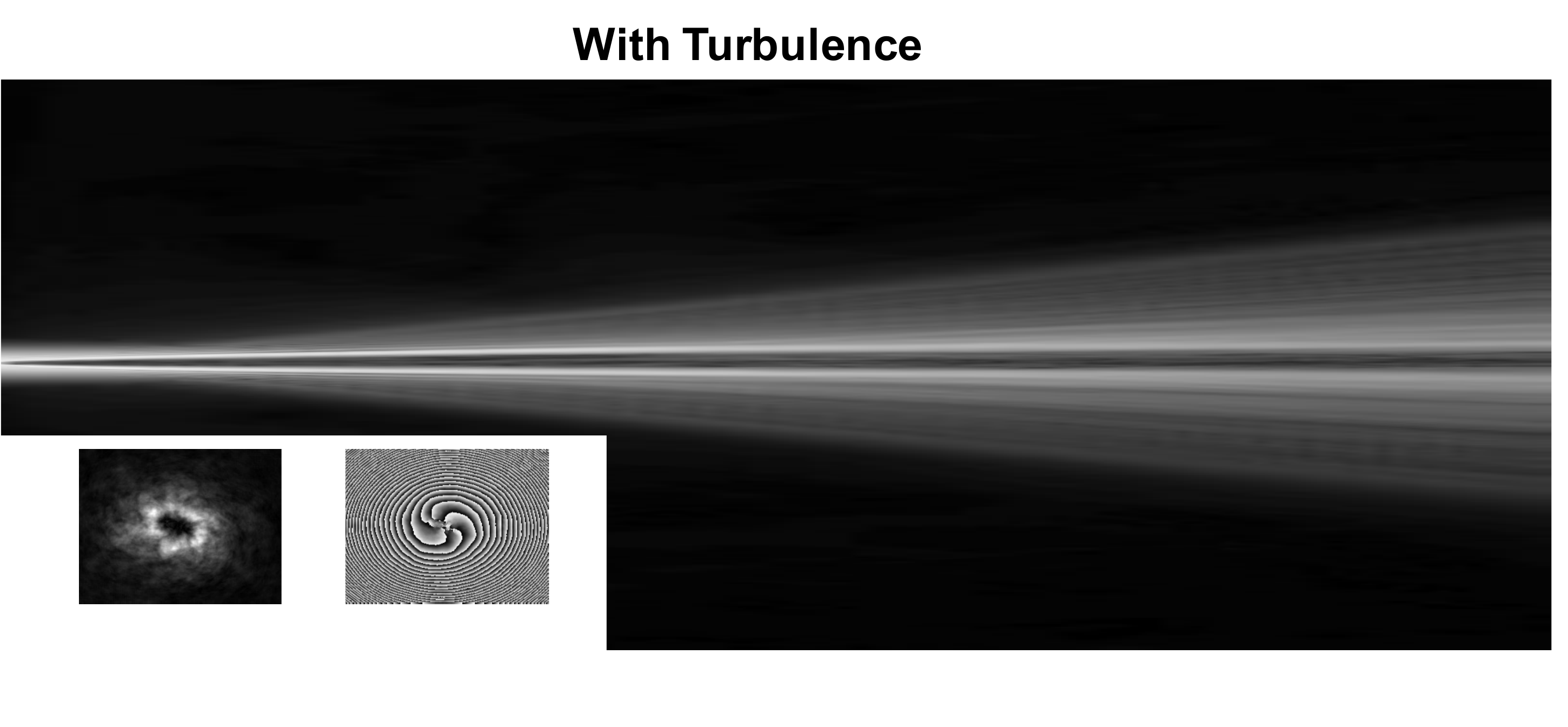}
  \end{tabular}  
  }
  \caption{(Top) Simulated propagation of a large number of rays through a pristine atmosphere.  The rays diverge linearly in time, consistent with our assumed action, Eqn. (\ref{eqn:action}). (Bottom) As the rays move through a turbulent atmosphere, simulated using 100 phase screens, they fluctuate slightly blurring the resulting image.  Nonetheless, the motion is still clearly dominated by kinetic energy with the variations in refractive index causing small changes to the motion.}
  \label{fig:justify}
\end{figure}
A numerical simulation of this method is shown in Figure (\ref{fig:justify}) in order to demonstrate how a ray-optics description of the EM field is influenced by the turbulence.  The upper plot shows a number of different optics rays propagating through a pristine (non-turbulent) atmosphere.  As expected the rays move in perfectly straight lines, thereby implying a constant velocity solution consistent with the action given by \eqref{eqn:action}.  The right plot shows the rays moving through a turbulent atmosphere as realized using 100 evenly spaced phase screens, designed to mimic the atmospheric properties of Kolmogorov turbulence.  While the rays clearly fluctuate over the path length, those fluctuations are minor relative to the main, linear trend.  Thus, we are capturing the turbulence-induced perturbations between the clean and corrupted image, but are modeling them as growing linearly over time in the transverse direction. Thus we conclude that, in an approximate sense, the deviations from a linear path are mostly local in time, in accordance with the result predicted from the optimal transport model expressed in Eqn. \eqref{eqn:interp2}. 

\subsection{Modeling turbulence in image time series}

\begin{figure}[tb]
    \center
    \includegraphics[scale=0.45]{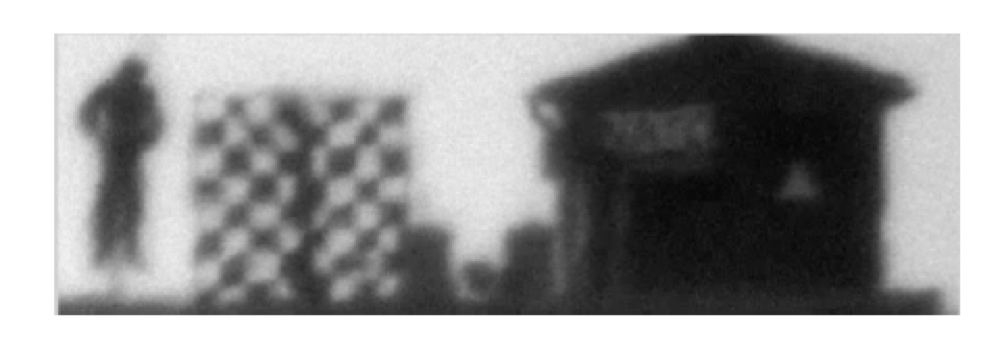} 
  
  \caption{Frame from video of a static scene imaged under turbulence due to atmospheric changes.}
  \label{fig:video_frame}
\end{figure}

In this section we analyze video data collected through a turbulent atmosphere and compare different modeling approaches with respect to their ability to describe the observed imagery. A frame from such video is shown in  Figure \ref{fig:video_frame}. The video shows a static scene, imaged through turbulent atmosphere, and thus contains the effects of noise, diffraction, and turbulence. As is commonly done, the models are compared in terms of 1) the error in the description and 2) the number of terms required of the description.  These are the two fundamental ingredients to all ``model selection'' methods we are aware of (see e.g., \cite{Burnham:98}). 

\

\begin{figure}[tb]
  \centerline{
    \includegraphics[scale=0.325]{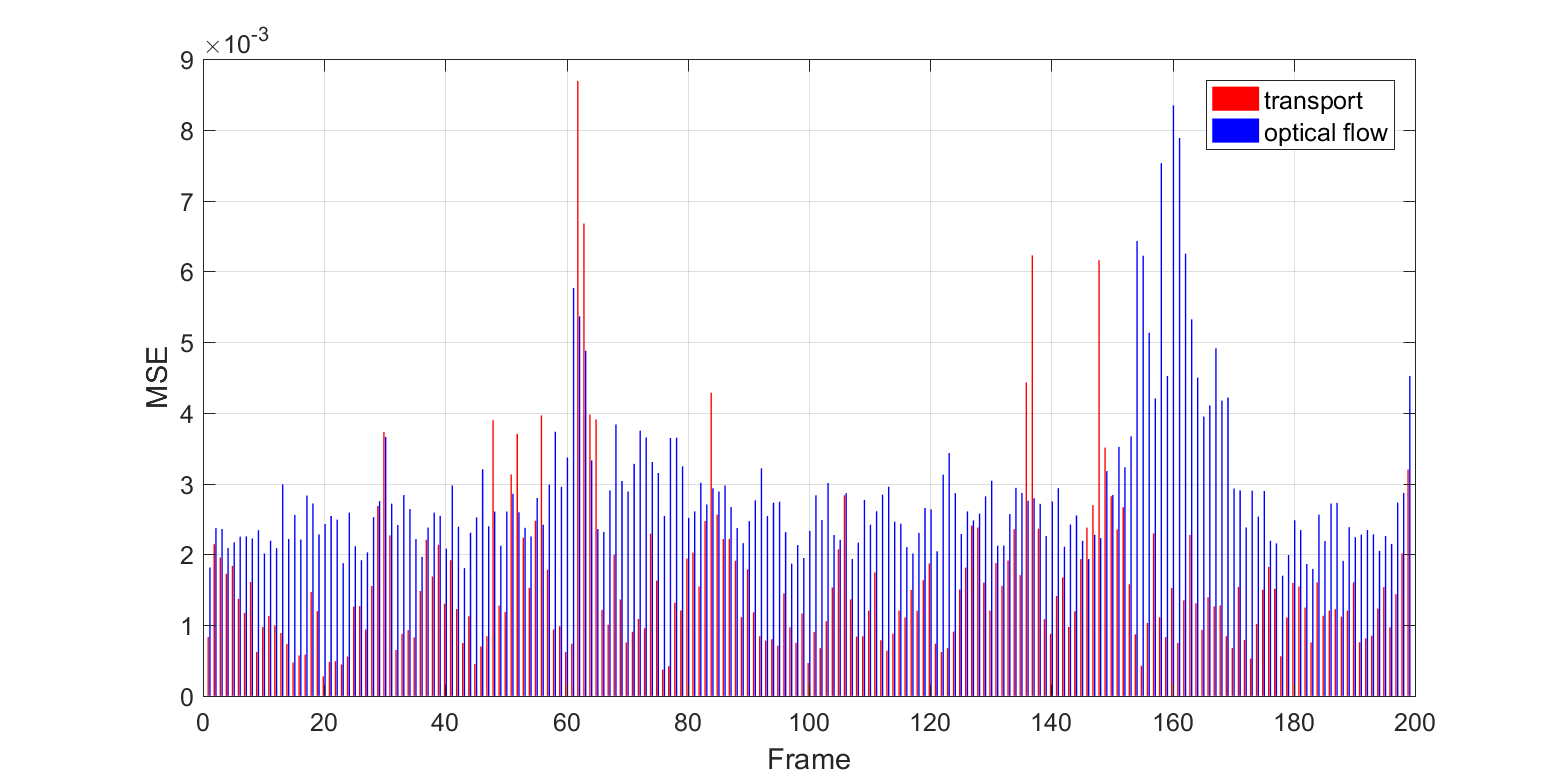} 
  }
  \caption{Comparison between the optical flow and transport models described in this section.  Shown are the mean-square error associated with frame-to-frame reconstruction showing that, as expected, the transport approach is able to obtain better matches between frames.}
  \label{fig:mse_result}
\end{figure}

Using the transport model described above, the underlying assumption is that (neglecting the effects of noise) the difference between two frames can be characterized by photon transport due to turbulence. Thus, from Eqn. \eqref{eqn:transform}, we hypothesize that $\det(J_f(\vec{x}_0,z))\rho(f(\vec{x}_0),z)=\rho(\vec{x}_0,0)$ where $\rho(\vec{x}_0,0)$ now represents the first frame of the movie, and $\rho(\vec{x}_0,z)$ is assumed to be the frame at time $t = z$. Taking the first frame as a reference, we seek to recover the information contained in the first frame from any other arbitrary frame using $f$ computed with an optimal transport code as described in \cite{Gustavo:16} that takes as input two images and outputs $f$ such that $\det(J_f(\vec{x}_0,z))\rho(f(\vec{x}_0),z)=\rho(\vec{x}_0,0)$ while simultaneously minimizing the action expressed in Eqn.  \eqref{eqn:Kanto}. For comparison purposes we also utilize an optical flow method \cite{Mitzel:09} for computing $g$ such that $\rho(g(\vec{x}_0),z) \sim \rho(\vec{x}_0,0)$, where the estimation is performed utilizing a regularized least squared error procedure. Results showing the mean squared error (MSE) between each reconstruction (using both transport $\det(J_f(\vec{x}_0,z))\rho(f(\vec{x}_0),z)$ and optical flow $\rho(g(\vec{x}_0),z)$ models) and the reference frame $\rho(\vec{x}_0,0)$ appear in Figure \ref{fig:mse_result}. 

The plot shows that the transport model is able to better match frames from the movie, which is an unsurprising result given that there exist multiple (infinite) $f$'s that will satisfy $\det(J_f(\vec{x}_0,z))\rho(f(\vec{x}_0),z)=\rho(\vec{x}_0,0)$ for any two normalized input images, while the same cannot be guaranteed for an optical flow (registration) model $\rho(g(\vec{x}_0),z) \sim \rho(\vec{x}_0,0)$.
 
\begin{figure}[tb]
  \centerline{
    \includegraphics[scale=0.5]{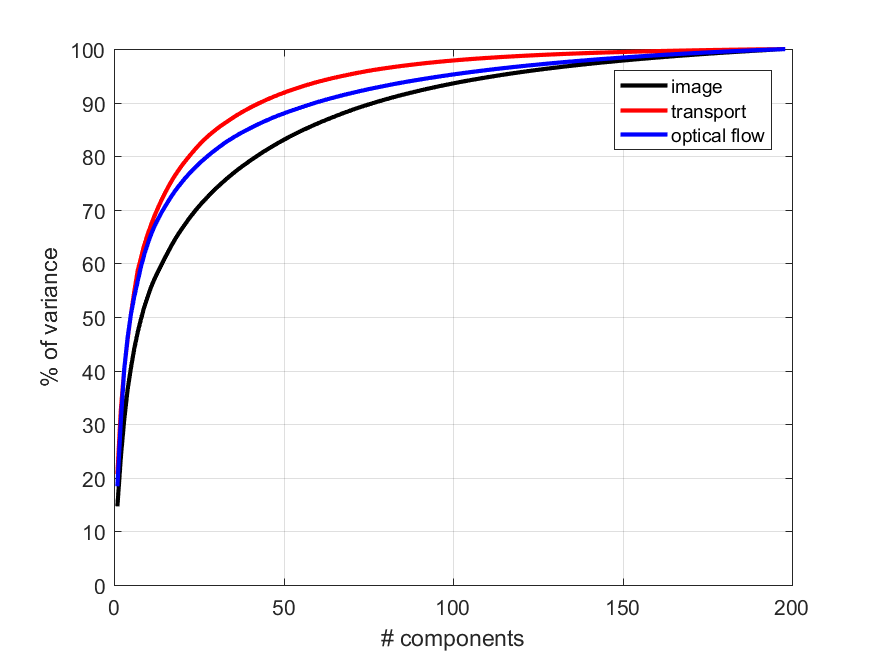} 
  }
  \caption{Percent of data set variance (normalized) as a function of the number of principal components used to model the input data in image space, optical flow, and transport models. The result shows that the transport model is the most parsimonious.}
  \label{fig:pca_result}
\end{figure}

We then sought to characterize the complexity present in the spatial transformation estimates computed via the transport and optical flow methods. Let $f_z$ correspond to the function that matches frame $z$ to frame $0$, that is $f_z$ is computed such that $\det(J_f(\vec{x}_0,z))\rho(f(\vec{x}_0),z)=\rho(\vec{x}_0,0)$. Similarly, we denote $g_z$ as the spatial transformation that matches $\rho(g(\vec{x}_0),z) \sim \rho(\vec{x}_0,0)$ using the optical flow model. Utilizing the standard principal component analysis (PCA) techniques we decompose the sequence of $f_z$, and respectively $g_z$, as a sum of eigen-functions (bases) computed using the PCA method. PCA is a technique that given a set of vectors, automatically discovers an ordered basis whereby the average MSE for reconstructing the dataset using only certain components (basis vectors or functions) is minimum. For comparison purposes, we also compute the eigen-decomposition of the image intensities for all frames (image space) as well. The percent of total variance captured as a function of the number of eigen-functions used in the reconstruction for all three spaces (transport, optical flow, and image) is shown in Fig. \ref{fig:pca_result} and shows that the transport model appears to be the most parsimonious model of all three.

\begin{figure}[tb]
  \centerline{
    \includegraphics[scale=0.5]{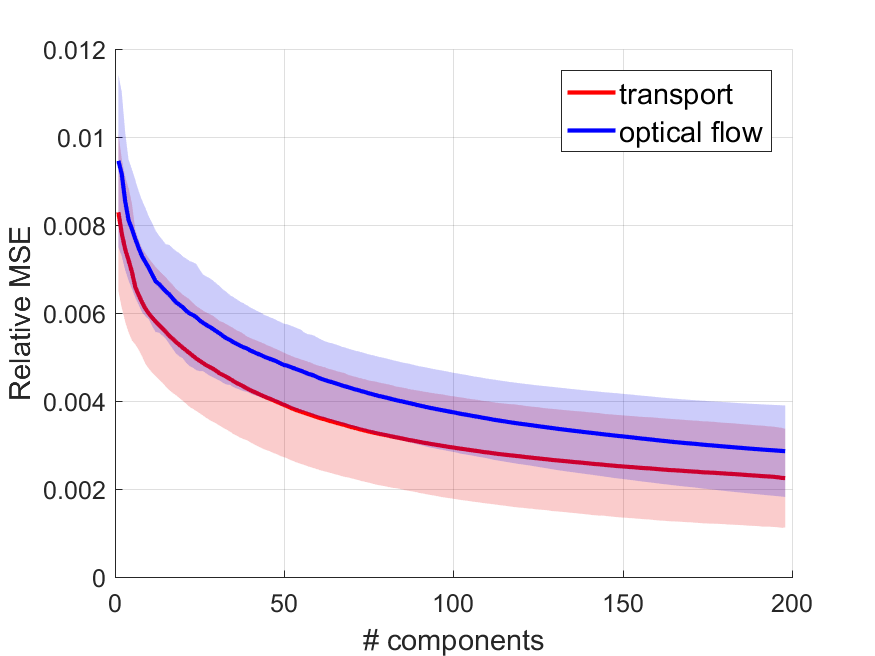} 
  }
  \caption{Mean square error of frame reconstruction of individual frames using both optical flow and transport models, as a function of the number of principal components used in each model, respectively.}
  \label{fig:mse_pca}
\end{figure}

Finally, we combine the MSE measurements described in Fig. \ref{fig:mse_result} with the PCA-derived parsimony measure displayed in Fig. \ref{fig:pca_result}. More specifically, here we investigate the ability of the PCA model for both transport and optical flow models to reconstruct the original frame $\rho(\vec{x}_0,0)$ as a function of the number of components utilized in estimating their respective transformations. Figure \ref{fig:mse_pca} shows the mean squared error between the original frame and the estimate of both transport and optical flow models, each using the same number of PCA components. In short, it is clear that for a fixed model complexity (a certain fixed number of basis functions used to model the transport or optical flows) the transport model more accurately reconstructs the original frame.

\section{Summary \& Discussion \label{sec:discussion}}

We have described a new approach for modeling the effects of turbulence in optical images using the principle of least action. In short, given only a pair of images (clean/corrupted) $\rho(\vec{x},0),~\rho(\vec{x},Z)$, and accepting the principle of least-action, we can solve Eqn. (\ref{eqn:Kanto}) and use the resulting map $f(\vec{x}_0,z)$ to obtain both the image intensity via (\ref{eqn:transform}) and phase function via (\ref{eqn:interp2}) at any point along the direction of propagation.   In doing so, we have effectively replaced explicit knowledge of the index fluctuations $\eta(\vec{x},z)$ with the physical principle of action minimization and a sample pair of images that have been so influenced.   We have further demonstrated that in solving (\ref{eqn:Kanto}) we are approximately solving parabolic wave equation for an image propagating in turbulent media, Eqn. (\ref{eqn:parabolic}).

The solution is exact as the propagation distance shrinks, or in the case that the refractive index does not possess a transverse gradient.  Given knowledge of the refractive index profile, however, one can augment the action (\ref{eqn:action}) and attempt to solve the system exactly, even in this more complicated situation.  Alternatively, given a sequence of images along the propagation path (see e.g., Fig. \ref{fig:path}), one could infer a piecewise-constant approximation of the refractive index profile.  Each of these extensions represents a potentially fruitful area of research.  

We believe the physical model described above could inform a new category of computational imaging methods for overcoming the barrier imposed by turbulence in open air imaging and communications. With regards to image enhancement, current algorithms for removing the effects of turbulence use an image registration-based procedure for spatially aligning (warping) sequential frames in a video segment \cite{Zhu:13,Furhad:16} . Our theory suggests that rather than being aligned, consecutive frames should be morphed instead via transport-based modeling. Moreover, the model linking clean and corrupted images should not be linear (e.g., ``deconvoution'' methods, see again \cite{Zhu:13,Furhad:16}), but should instead involve the inversion of optimal transport.
	
In yet another application, orbital angular momentum has recently been used to develop free-space optical communication strategies that augment the throughput of existing links \cite{Willner:15}. State of the art methods for decoding the nonlinear effects of turbulent channels involve the use of deep convolutional neural networks \cite{Doster:16}, and hence have a limited bandwidth (e.g. $\sim 1$ kilo bits/second) due to the high computational cost. The modeling described above can potentially be used to inform more computationally efficient decoding methods.

%

\appendices

\section{Continuity and Momentum from the Parabolic Wave Equation}\label{sec:appendixA}
The parabolic wave equation is written \cite{Fannjiang:05}
\begin{align}
i2k_0\partial_z \Psi(\vec{x},z)+\nabla_X^2 \Psi(\vec{x},z)+k_0^2\eta(\vec{x},z)\Psi(\vec{x},z)=0
\end{align}
where $k_0$ is the wavenumber, $\eta(\vec{x},z)$ is the perturbation to the refractive index, i.e., $n^2(\vec{x},z)=1+\eta(\vec{x},z)$.  The EM field $\Psi(\vec{x},z)$ is in $V/m$ and the notation $\nabla_X^2=\partial^2/\partial_{x_1}^2+\partial^2/\partial_{x_2}^2$ is the Laplacian w.r.t. the transverse coordinates $\vec{x}\equiv (x_1,x_2)$ and $z$ is the direction of propagation.  Henceforth we will remove the arguments and simply note that the EM field, magnitude, and phase are all functions of the transverse coordinates $\vec{x}$ and $z$.  Now rescale the spatial coordinates by the wavelength so that $z'=\frac{k_0}{2}z$, $x_1'=k_0x_1$ and $x_2'=k_0x_2$ in which case the spatially non-dimensionalized wave equation becomes
\begin{align}
i\partial_{z'} \Psi+\nabla_{X'}^2 \Psi+\eta\Psi=0.
\label{eqn:parabolicA}
\end{align}
To transform this expression we can use the so-called Madelung transformation which sets $\Psi(\vec{x}',z')\equiv \rho(\vec{x}',z')^{1/2}e^{i\phi(\vec{x}',z')/2}$.  

For ease of notation we drop the $'$ and state explicitly that we are working with non-dimensional lengths.  Form the identity
\begin{align}
\frac{\nabla_X \Psi}{\Psi}&=\frac{\frac{1}{2}\rho^{-1/2}\nabla_X\rho e^{i\phi/2}+i\frac{1}{2}e^{i\phi/2}\rho^{1/2}\nabla_X\phi}{\rho^{1/2}e^{i\phi/2}}\nonumber \\
&=\frac{\nabla_X\rho}{2\rho}+i\frac{1}{2}\nabla_X\phi.
\label{eqn:ident1}
\end{align}
Recognizing that $\rho=\Psi\Psi^*$ and substituting into (\ref{eqn:ident1}) 
\begin{align}
\frac{\nabla_X\Psi}{\Psi}&=\frac{\nabla_X\left(\Psi\Psi^*\right)}{2\Psi\Psi^*}+\frac{i}{2}\nabla_X\phi\nonumber \\
&=\frac{(\nabla_X\Psi)\Psi^*+(\nabla_X\Psi^*)\Psi}{2\Psi\Psi^*}+\frac{i}{2}\nabla_X\phi\nonumber \\
\frac{\nabla_X\Psi}{2\Psi}-\frac{\nabla_X\Psi^*}{2\Psi^*}&=\frac{i}{2}\nabla_X\phi\nonumber
\end{align}
and then multiplying both sides by $\rho=\Psi\Psi^*$
\begin{align}
(\nabla_X\Psi)\Psi^*-(\nabla_X\Psi^*)\Psi^*&=i\rho\nabla_X\phi\nonumber
\end{align}
and finally taking the divergence of both sides gives
\begin{align}
(\nabla^2_X\Psi)\Psi^*&+\nabla_X\Psi^*\nabla_X\Psi-(\nabla^2_X\Psi^*)\Psi-\nabla_X\Psi\nabla_X\Psi^*\nonumber \\
&=i\nabla_X\cdot(\rho\nabla_X\phi)\nonumber \\
\nabla_X\cdot(\rho\nabla_X\phi)&=-i[(\nabla^2_X\Psi)\Psi^*-(\nabla^2_X\Psi^*)\Psi].
\label{eqn:ident2}
\end{align}
Now, returning to (\ref{eqn:parabolicA}) we note that the complex conjugate of the EM field similarly satisfies
\begin{align}
-i\partial_{z} \Psi^*+\nabla_X^2 \Psi^*+\eta\Psi^*=0.
\label{eqn:parabolicB}
\end{align}
Multiplying (\ref{eqn:parabolicA}) by $-i\Psi^*$ and (\ref{eqn:parabolicB}) by $i\Psi$ and adding gives
\begin{align}
&\left(\partial_{z} \Psi\right)\Psi^*-i(\nabla_X^2 \Psi)\Psi^*-i\rho \eta=0\nonumber \\
+~~~&\left(\partial_{z} \Psi^*\right)\Psi+i(\nabla_X^2 \Psi^*)\Psi+i\rho \eta=0.\nonumber \\
\hline\nonumber \\
&\partial_{z}\rho-i[(\nabla^2_X\Psi)\Psi^*-(\nabla^2_X\Psi^*)\Psi]=0\nonumber 
\end{align}
which can be combined with (\ref{eqn:ident2}) to yield
\begin{align}
\partial_{z}\rho+\nabla_X\cdot(\rho\nabla_X\phi)=0
\end{align}
which, after defining $v=\nabla_X\phi$, is exactly the continuity equation.  Note that the velocity is dimensionless as are the distances associated with differentiation.  The units are thus dictated solely by the units of $\rho$ which are $V^2/m^2$.

To obtain the momentum equation one again uses the identities $\Psi=\rho^{1/2}e^{i\phi/2}$, $\rho=\Psi\Psi^*$, $v=\nabla_X\phi$ and substitute directly into (\ref{eqn:parabolic}).  Making note of the previous result (the terms of the continuity equation appear and can therefore be set equal to zero), and using the identity
\begin{align}
\frac{\nabla_X^2\rho}{2\rho}-\frac{(\nabla_X\rho)^2}{4\rho^2}=\frac{\nabla_X^2(\rho^{1/2})}{\rho^{1/2}}
\end{align}
one has
\begin{align}
\partial_z\phi+\frac{1}{2}(\nabla_X\phi)^2=2\frac{\nabla_X^2(\rho^{1/2})}{\rho^{1/2}}+2\eta.
\label{eqn:phase5}
\end{align}
The term $\frac{\nabla_X^2(\rho^{1/2})}{\rho^{1/2}}$ is referred to in optics as the ``diffraction term'' \cite{Gureyev:95}, or in the quantum literature, the ``quantum potential'' \cite{Bohm:84}.  This term is typically neglected in optics given certain assumptions about the spatial variability in intensity with respect to a wavelength \cite{Saleh:91},~\cite{Gureyev:95}.  We too neglect this term, and in the next section (\ref{sec:appendixB}) provide additional justification for its removal. 

Finally, taking the spatial gradient $\nabla_X$ of both sides and recognize that $\nabla_X\left[(\nabla_X\phi)^2\right]=\nabla_X(v\cdot v)=2(v\times(\nabla_X\times v))+2(v\cdot\nabla_X)v$.  Noting that a necessary and sufficient condition for representing the velocity as the gradient of a potential is $\nabla_X\times v=0$ \cite{Panton:84} we finally obtain the form (\ref{eqn:momentum2}).

\section*{Simplification of the diffraction term }\label{sec:appendixB}

In the derivation of the wave equation we excluded the divergence of the electric field on the physical reasoning that the fluctuations in the atmosphere were relatively minor.  In what follows, however, we show that the constitutive law given by \eqref{eqn:constraint} can be used to relate the diffraction term in \eqref{eqn:phase5} to the refractive index and, by extension, to better understand the conditions under which this term can be safely neglected.

Returning to the vector description of the electric field, for linearly polarized light we may write $\vec{E}(\vec{x})=\{\rho^{1/2}\cos(\gamma)\hat{x}_1,~\rho^{1/2}\sin(\gamma)\hat{x}_2\}$ where $\gamma$ is the polarization angle, measured relative to the $\hat{x}_1$.

Using this representation for the electric field we can expand the relationship expressed in (\ref{eqn:constraint}) as
\begin{align}
\nabla_X\cdot &\rho^{1/2}\left[\cos\left(\gamma\right)\hat{x}_1+\sin\left(\gamma\right)\hat{x}_2\right]=\nonumber \\
&-\rho^{1/2}\left[\cos\left(\gamma\right)\hat{x}_1+\sin\left(\gamma\right)\hat{x}_2\right]\cdot 2\frac{\nabla_X n}{n}.
\label{eqn:test5}
\end{align}
Expanding the first line gives
\small
\begin{align}
&\nabla_X\cdot \rho^{1/2}\left[\cos\left(\gamma\right)\hat{x}_1+\sin\left(\gamma\right)\hat{x}_2\right]=\nonumber \\
&~~\nabla_X \rho^{1/2}\cdot \left[\cos\left(\gamma\right)\hat{x}_1+\sin\left(\gamma\right)\hat{x}_2\right]\nonumber \\
&~~+\rho^{1/2}\left[-\frac{\partial \gamma}{\partial x_1}\hat{x}_1+\frac{\partial \gamma}{\partial x_2}\hat{x}_2\right]\cdot \left[\sin\left(\gamma\right)\hat{x}_1+\cos\left(\gamma\right)\hat{x}_2\right]
\end{align}
\normalsize
so that the entire expression given by \eqref{eqn:test5} can be written
\begin{align}
&\left\{\frac{\nabla_X\rho^{1/2}}{\rho^{1/2}}+2\frac{\nabla_X n}{n}\right\}\cdot \left[\cos\left(\gamma\right)\hat{x}_1,~\sin\left(\gamma\right)\hat{x}_2\right]\nonumber \\
&+\nabla_X\times \left[-\sin\left(\gamma\right)\hat{x}_1,~\cos\left(\gamma\right)\hat{x}_2\right]=0.
\end{align}
For the expression to hold for arbitrary angle of polarization (which may be different at every spatial location $\vec{x}$ \cite{Born:99}), the term in brackets must equate to zero.  Thus, simplifying the intensity term and rearranging we have
\begin{align}
\frac{\nabla_X \rho}{2\rho}&=-2\frac{\nabla_X n}{n}
\label{eqn:constitagain}
\end{align}
This is a vector equation relating intensity and polarization angle to the refractive index in the transverse plane.  The term involving the curl of $ \left[-\sin\left(\gamma\right)\hat{x}_1,~\cos\left(\gamma\right)\hat{x}_2\right]$ points in the direction of propagation hence it can be set equal to zero.

Now, taking the divergence of both sides of the remaining terms in (\ref{eqn:constitagain}) gives
\begin{align}
\nabla_X\cdot \frac{\nabla_X \rho}{2\rho}&=-2\nabla_X\cdot \frac{\nabla_X n}{n}.
\end{align}

Continuing with the divergence operator we have
\begin{align}
\frac{\nabla_X^2\rho}{2\rho}&-\frac{(\nabla_X \rho)^2}{2\rho^2}=-2\nabla_X\cdot \left(\frac{\nabla_X n}{n}\right)\nonumber \\
\frac{\nabla_X^2\rho}{2\rho}&-\frac{(\nabla_X \rho)^2}{2\rho^2}=-2\left[\frac{\nabla_X^2n}{n}-\frac{(\nabla_X n)^2}{n^2}\right]
\label{eqn:constit3}
\end{align}
The term on the left hand side can be split into three terms, two of which we already know how to combine into what we need.  Specifically,
\begin{align}
\frac{\nabla_X^2\rho}{2\rho}-\frac{(\nabla_X \rho)^2}{2\rho^2}&=\frac{\nabla_X^2\rho}{2\rho}-\frac{(\nabla_X \rho)^2}{4\rho^2}
-\frac{(\nabla_X \rho)^2}{4\rho^2}\nonumber \\
&=\frac{\nabla_X^2\rho^{1/2}}{\rho^{1/2}}-\frac{(\nabla_X \rho)^2}{4\rho^2}
\end{align}
in which case \eqref{eqn:constit3} becomes
\begin{align}
\frac{\nabla_X^2\rho^{1/2}}{\rho^{1/2}}&=\frac{(\nabla_X \rho)^2}{4\rho^2}-2\left[\frac{\nabla_X^2n}{n}-\frac{(\nabla_X n)^2}{n^2}\right]
\label{eqn:constit4}
\end{align}
However, by squaring both sides of \eqref{eqn:constitagain} we can replace the first term on the right-hand-side of \eqref{eqn:constit4} so that
\begin{align}
&\frac{\nabla^2\rho^{1/2}}{\rho^{1/2}}=-2\frac{\nabla_X^2n}{n}+6\left(\frac{\nabla_Xn}{n}\right)^2\nonumber \\
&=-2\left[\nabla\cdot \frac{\nabla_Xn}{n}+\left(\frac{\nabla_Xn}{n}\right)^2\right]+6\left(\frac{\nabla_Xn}{n}\right)^2\nonumber \\
&=-2\nabla\cdot \frac{\nabla_Xn}{n}+4\left(\frac{\nabla_Xn}{n}\right)^2\nonumber \\
&=-\nabla^2_X\log(n^2)+(\nabla_X\log(n^2))^2
\end{align}
Thus, for small perturbations to the index $\eta\ll 1$ the approximation $\log(1+\delta)\approx \delta$ for $\delta\ll 1$ means we could alternatively have written the last line above as $-\nabla^2_X\eta+(\nabla_X\eta)^2$.  These terms are clearly higher-order in terms of the index perturbations, hence are properly neglected in the analysis.

\bibliographystyle{IEEEtran}
\bibliography{refs}
\end{document}